\documentclass[12pt]{article}
\usepackage{graphicx}

\usepackage[absolute,overlay]{textpos}
\usepackage{amsmath}
\usepackage{caption}

\usepackage{graphicx}
\usepackage{subfloat}
\usepackage{subfig}

\usepackage{lmodern}
\usepackage{multirow}
\usepackage{mathtools}  
\mathtoolsset{showonlyrefs}
\usepackage{xspace}
\usepackage{caption}

\makeatletter
\newcommand*{\rom}[1]{\expandafter\@slowromancap\romannumeral #1@}
\makeatother
\usepackage{amssymb}
\usepackage{amsfonts}
\usepackage{upgreek}

\usepackage{xspace} 
                    
\newcommand*\diff{\mathop{}\!\mathrm{d}}

\usepackage{color}
\usepackage{colortbl}
\graphicspath{{./figs/}} 

\usepackage{amsmath} 
\usepackage{amssymb}
\usepackage{amsfonts}
\usepackage{upgreek} 

\usepackage{ifthen} 
\newboolean{uprightparticles}
\usepackage{upgreek}

\def\lhcb {\mbox{LHCb}\xspace}

\def\babar  {\mbox{BaBar}\xspace}
\def\belle  {\mbox{Belle}\xspace}

\def\lhc    {\mbox{LHC}\xspace}

\def\rich   {RICH\xspace}

\def\hlt    {HLT\xspace}
\def\hltone {HLT1\xspace}
\def\hlttwo {HLT2\xspace}

\ifthenelse{\boolean{uprightparticles}}%
{

 \def\Peta        {\ensuremath{\upeta}\xspace}

 \def\Pmu         {\ensuremath{\upmu}\xspace}

 \def\Ppi         {\ensuremath{\uppi}\xspace}

 \def\Ptau        {\ensuremath{\uptau}\xspace}

 \def\PDelta      {\ensuremath{\Delta}\xspace}                 
 \def\PXi      {\ensuremath{\Xi}\xspace}                 
 \def\PLambda      {\ensuremath{\Lambda}\xspace}                 
 \def\PSigma      {\ensuremath{\Sigma}\xspace}                 
 \def\POmega      {\ensuremath{\Omega}\xspace}                 
 \def\PUpsilon      {\ensuremath{\Upsilon}\xspace}                 
 
 \def\PB      {\ensuremath{\mathrm{B}}\xspace}                 
                  
 \def\PD      {\ensuremath{\mathrm{D}}\xspace}

 \def\PK      {\ensuremath{\mathrm{K}}\xspace}

 \def\PR      {\ensuremath{\mathrm{R}}\xspace}                 
 \def\PS      {\ensuremath{\mathrm{S}}\xspace}

 \def\Pb      {\ensuremath{\mathrm{b}}\xspace}                 
 \def\Pc      {\ensuremath{\mathrm{c}}\xspace}                 
 \def\Pd      {\ensuremath{\mathrm{d}}\xspace}

 \def\Pi      {\ensuremath{\mathrm{i}}\xspace}

 \def\Pp      {\ensuremath{\mathrm{p}}\xspace}

 \def\Ps      {\ensuremath{\mathrm{s}}\xspace}                 
                  
 \def\Pu      {\ensuremath{\mathrm{u}}\xspace}

}
{

 \def\Peta        {\ensuremath{\eta}\xspace}

 \def\Pmu         {\ensuremath{\mu}\xspace}

 \def\Ppi         {\ensuremath{\pi}\xspace}

 \def\Ptau        {\ensuremath{\tau}\xspace}

 \mathchardef\PDelta="7101
 \mathchardef\PXi="7104
 \mathchardef\PLambda="7103
 \mathchardef\PSigma="7106
 \mathchardef\POmega="710A
 \mathchardef\PUpsilon="7107
                  
 \def\PB      {\ensuremath{B}\xspace}                 
                  
 \def\PD      {\ensuremath{D}\xspace}

 \def\PK      {\ensuremath{K}\xspace}

 \def\PR      {\ensuremath{R}\xspace}                 
 \def\PS      {\ensuremath{S}\xspace}

 \def\Pb      {\ensuremath{b}\xspace}                 
 \def\Pc      {\ensuremath{c}\xspace}                 
 \def\Pd      {\ensuremath{d}\xspace}

 \def\Pi      {\ensuremath{i}\xspace}

 \def\Pp      {\ensuremath{p}\xspace}

 \def\Ps      {\ensuremath{s}\xspace}                 
                  
 \def\Pu      {\ensuremath{u}\xspace}

 \def\Ppm     {\ensuremath{\pm}\xspace}                 
}

\DeclareRobustCommand{\optbar}[1]{\shortstack{{\miniscule (\rule[.5ex]{1.25em}{.18mm})}
  \\ [-.7ex] $#1$}}

\def\mup        {{\ensuremath{\Pmu^+}}\xspace}
\def\mun        {{\ensuremath{\Pmu^-}}\xspace} 

\def\uquark    {{\ensuremath{\Pu}}\xspace}

\def\dquark    {{\ensuremath{\Pd}}\xspace}

\def\squark    {{\ensuremath{\Ps}}\xspace}

\def\cquark    {{\ensuremath{\Pc}}\xspace}
\def\cquarkbar {{\ensuremath{\overline \cquark}}\xspace}
\def\ccbar     {{\ensuremath{\cquark\cquarkbar}}\xspace}
\def\bquark    {{\ensuremath{\Pb}}\xspace}

\def\pion   {{\ensuremath{\Ppi}}\xspace}

\def\pip    {{\ensuremath{\pion^+}}\xspace}

\def\kaon    {{\ensuremath{\PK}}\xspace}
  \def\Kbar    {{\kern 0.2em\overline{\kern -0.2em \PK}{}}\xspace}

\def\KorKbar    {\kern 0.18em\optbar{\kern -0.18em K}{}\xspace}

\def\Km      {{\ensuremath{\kaon^-}}\xspace}

  \def\Dbar    {{\kern 0.2em\overline{\kern -0.2em \PD}{}}\xspace}
\def\D       {{\ensuremath{\PD}}\xspace}

\def\DorDbar    {\kern 0.18em\optbar{\kern -0.18em D}{}\xspace}

\def\Dp      {{\ensuremath{\D^+}}\xspace}

\def\B       {{\ensuremath{\PB}}\xspace}
\def\Bbar    {{\ensuremath{\kern 0.18em\overline{\kern -0.18em \PB}{}}}\xspace}

\def\BorBbar    {\kern 0.18em\optbar{\kern -0.18em B}{}\xspace}

  \def\Y#1S{\ensuremath{\PUpsilon{(#1S)}}\xspace}

\def\proton      {{\ensuremath{\Pp}}\xspace}

\def\Xires       {{\ensuremath{\PXi}}\xspace}

\def\Lz          {{\ensuremath{\PLambda^0}}\xspace}
\def\Lbar        {{\ensuremath{\kern 0.1em\overline{\kern -0.1em\PLambda}}}\xspace}
\def\LorLbar    {\kern 0.18em\optbar{\kern -0.18em \PLambda}{}\xspace}
\def\Lambdares   {{\ensuremath{\PLambda}}\xspace}

\def\Sigmares    {{\ensuremath{\PSigma}}\xspace}

\def\Omegares    {{\ensuremath{\POmega}}\xspace}

\def\Lc      {{\ensuremath{\Lz^+_\cquark}}\xspace}

\def\Xicz    {{\ensuremath{\Xires^0_\cquark}}\xspace}
\def\Xicp    {{\ensuremath{\Xires^+_\cquark}}\xspace}

\def\BF         {{\ensuremath{\cal B}}\xspace}

\newcommand{\decay}[2]{\ensuremath{#1\!\to #2}\xspace}         

\def\to                 {\ensuremath{\rightarrow}\xspace}

\def\CP                {{\ensuremath{C\!P}}\xspace}

\newcommand{\dm}{{\ensuremath{\delta m}}\xspace}

\def\AT#1     {\ensuremath{A_{\mathrm{T}}^{#1}}\xspace}           

\def\C#1      {\ensuremath{\mathcal{C}_{#1}}\xspace}                       
\def\Cp#1     {\ensuremath{\mathcal{C}_{#1}^{'}}\xspace}                    
\def\Ceff#1   {\ensuremath{\mathcal{C}_{#1}^{\mathrm{(eff)}}}\xspace}        
\def\Cpeff#1  {\ensuremath{\mathcal{C}_{#1}^{'\mathrm{(eff)}}}\xspace}       
\def\Ope#1    {\ensuremath{\mathcal{O}_{#1}}\xspace}                       
\def\Opep#1   {\ensuremath{\mathcal{O}_{#1}^{'}}\xspace}                    

\newcommand{\tev}{\ensuremath{\mathrm{\,Te\kern -0.1em V}}\xspace}
\newcommand{\gev}{\ensuremath{\mathrm{\,Ge\kern -0.1em V}}\xspace}
\newcommand{\mev}{\ensuremath{\mathrm{\,Me\kern -0.1em V}}\xspace}
\newcommand{\kev}{\ensuremath{\mathrm{\,ke\kern -0.1em V}}\xspace}
\newcommand{\ev}{\ensuremath{\mathrm{\,e\kern -0.1em V}}\xspace}
\newcommand{\gevc}{\ensuremath{{\mathrm{\,Ge\kern -0.1em V\!/}c}}\xspace}
\newcommand{\mevc}{\ensuremath{{\mathrm{\,Me\kern -0.1em V\!/}c}}\xspace}
\newcommand{\gevcc}{\ensuremath{{\mathrm{\,Ge\kern -0.1em V\!/}c^2}}\xspace}
\newcommand{\gevgevcccc}{\ensuremath{{\mathrm{\,Ge\kern -0.1em V^2\!/}c^4}}\xspace}
\newcommand{\mevcc}{\ensuremath{{\mathrm{\,Me\kern -0.1em V\!/}c^2}}\xspace}

\newcommand{\mytev}{\ensuremath{\mathrm{\,Te\kern -0.1em V}}\xspace}

\def\cm   {\ensuremath{\rm \,cm}\xspace}

\def\mm   {\ensuremath{\rm \,mm}\xspace}

\def\mub{\ensuremath{{\rm \,\upmu b}}\xspace}

\def\invnb {\ensuremath{\mbox{\,nb}^{-1}}\xspace}

\def\invpb {\ensuremath{\mbox{\,pb}^{-1}}\xspace}

\def\invfb   {\ensuremath{\mbox{\,fb}^{-1}}\xspace}

\def\fs   {\ensuremath{\rm \,fs}\xspace}

\def\khz  {\ensuremath{{\rm \,kHz}}\xspace}

\newcommand{\stat}{\ensuremath{\mathrm{\,(stat)}}\xspace}
\newcommand{\syst}{\ensuremath{\mathrm{\,(syst)}}\xspace}

\newcommand{\chisq}{\ensuremath{\chi^2}\xspace}

\def\gsim{{~\raise.15em\hbox{$>$}\kern-.85em
          \lower.35em\hbox{$\sim$}~}\xspace}
\def\lsim{{~\raise.15em\hbox{$<$}\kern-.85em
          \lower.35em\hbox{$\sim$}~}\xspace}

\def\sqs   {\ensuremath{\protect\sqrt{s}}\xspace}

\def\pt         {\mbox{$p_{\rm T}$}\xspace}

\newcommand{\lum} {\ensuremath{\mathcal{L}}\xspace}
\newcommand{\lumi} {\ensuremath{\mathcal{L}_{\mathrm{int}}}\xspace}

\def\pythia     {\mbox{\textsc{Pythia}}\xspace}

\def\tell1  {TELL1\xspace}
\def\ukl1   {UKL1\xspace}

\def\Lc      {\ensuremath{\Lambdares_\cquark}\xspace}

\def\Lcp     {\ensuremath{\Lambdares_\cquark^+}\xspace}

\renewcommand{\decay}[2]{\mbox{\ensuremath{#1\!\to #2}}\xspace}   

\newcommand{\TeV}{\tev}

\newcommand{\keV}{\keV}

\newcommand{\MeVcc}{\mevcc}

\def\pT         {\pt}

\def\Xiccp {\ensuremath{\Xi_{cc}^+}\xspace}

\newcommand{\LcpTopKmpip}{\decay{\Lcp}{\proton \Km \pip}}

\newcommand{\ip}{\ensuremath{\mathit{IP}}\xspace}	

\newcommand{\logipchisq}{\ensuremath{\log(\ip\chisq})\xspace}

\newcommand{\Tabref}[1]{Table~\protect\ref{#1}}

\newcommand{\Figref}[1]{Figure~\protect\ref{#1}}

\newcommand{\bi}{\begin{itemize}}
\newcommand{\ei}{\end{itemize}}

\newcommand{\bif}{ \begin{itemize} \footnotesize} 
\newcommand{\bis}{ \begin{itemize} \scriptsize}

\newcommand{\runone}{Run \rom{1}\xspace}
\newcommand{\runtwo}{Run \rom{2}\xspace}
\newcommand{\runthree}{Run \rom{3}\xspace}

\def\Xiccp {\ensuremath{\Xi_{\cquark\cquark}^{+}}\xspace}

\def\Sigmacpp {\ensuremath{\Sigmares_{\cquark}^{++}}\xspace}
\def\Sigmacz  {\ensuremath{\Sigmares_{\cquark}^{0}}\xspace}

\def\Omegaccc{\ensuremath{\Omega_{\cquark\cquark\cquark}^{+++}}\xspace}

\newcommand{\sufour}{\mbox{\ensuremath{SU(4)}}\xspace}

\def\Lc      {{\ensuremath{\Lz^+_\cquark}}\xspace}

\def\Xicz    {{\ensuremath{\Xires^0_\cquark}}\xspace}
\def\Xicp    {{\ensuremath{\Xires^+_\cquark}}\xspace}
\def\Xiccp    {{\ensuremath{\Xires^+_{\cquark\cquark}}}\xspace}
\def\Xiccpp    {{\ensuremath{\Xires^{++}_{\cquark\cquark}}}\xspace}

\def\Omegacc    {{\ensuremath{\Omegares^+_{\cquark\cquark}}}\xspace}
\def\Omegaccc    {{\ensuremath{\Omegares^+_{\cquark\cquark\cquark}}}\xspace}

\newboolean{inbibliography}
\setboolean{inbibliography}{false}
\newcommand\pubnumber{WSU--HEP--XXYY}
\newcommand\pubdate{\today}

\def\wayne{Department of Physics and Astronomy\\
University of Glasgow, Glasgow, UK}
\def\support{\footnote{on behalf of the \lhcb collaboration.}}

\def\Title#1{\begin{center} {\Large #1 } \end{center}}
\def\Author#1{\begin{center}{ \sc #1} \end{center}}
\def\Address#1{\begin{center}{ \it #1} \end{center}}

\newcommand\pubblock{\rightline{\begin{tabular}{l} \pubnumber\\
         \pubdate  \end{tabular}}}
\newenvironment{Abstract}{\begin{quotation}  }{\end{quotation}}
\newenvironment{Presented}{\begin{quotation} \begin{center} 
             PRESENTED AT\end{center}\bigskip 
      \begin{center}\begin{large}}{\end{large}\end{center} \end{quotation}}
\def\Acknowledgements{\bigskip  \bigskip \begin{center} \begin{large}
             \bf ACKNOWLEDGEMENTS \end{large}\end{center}}

\begin{document}
\begin{titlepage}
\pubblock

\vfill
\Title{Charmed baryons from \lhcb}
\vfill
\Author{Stephen Ogilvy\support}
\Address{\wayne}
\vfill
\begin{Abstract}
The vast amount of \ccbar production that can be recorded by the \lhcb detector makes it an ideal environment to study the hadronic production of charmed baryons, along with the properties of their decays.
We briefly describe the \lhcb experiment and the triggering mechanisms it uses for recording charm production. Previous charmed baryon results from \lhcb are detailed, with a description of the future plans for the charmed baryon programme.
\end{Abstract}
\vfill
\begin{Presented}
The 7th International Workshop on Charm Physics (CHARM 2015)\\
Detroit, MI, 18-22 May, 2015
\end{Presented}
\vfill
\end{titlepage}
\def\thefootnote{\fnsymbol{footnote}}
\setcounter{footnote}{0}
%

\section{Introduction}
\begin{figure}[htb]
\centering
\includegraphics[height=4.0in]{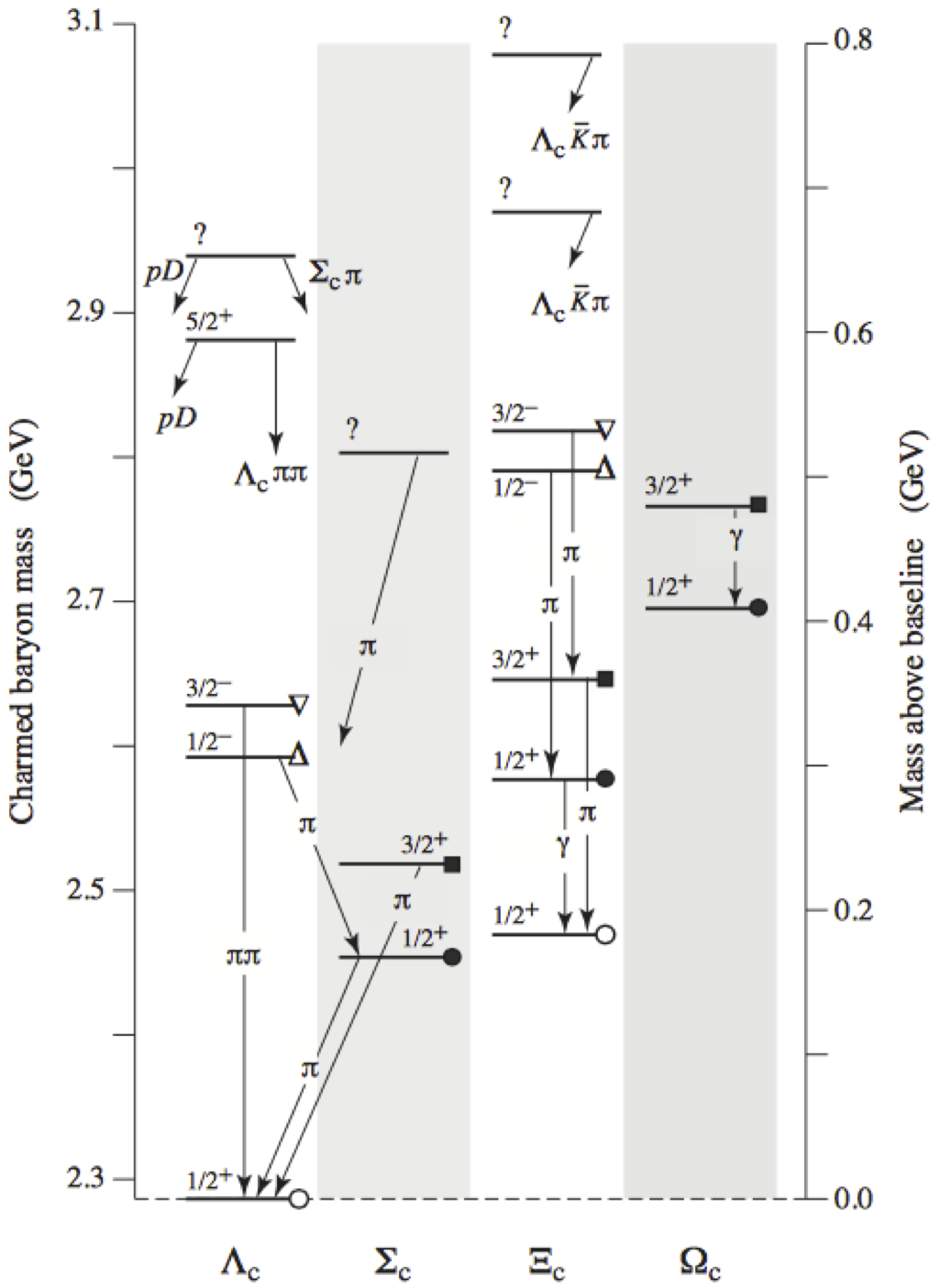}
\caption{The spectra of known singly-charmed baryons and their mass splittings.
Reproduced from~\cite{PDG2012}.
}
\label{fig:singlecharm}
\end{figure}

Recent years have seen great advances in our understandings of the charmed baryons.
The \B-factories, in particular the \belle and \babar experiments, have been successful in making a wide variety of first observations of excited singly-charmed baryons~\cite{Chistov:2006zj, Lesiak:2006sk, Mizuk:2004yu, Aubert:2007dt, Lesiak:2008wz, Aubert:2007bt, Aubert:2006je, Solovieva:2008fw}.
The current best knowledge of the spectra of singly-charmed baryons is given in \Figref{fig:singlecharm}.
The spin-parity assignments of many of the observed states are still to be discovered.

In this document we briefly describe the \lhcb detector and the strategies used in the triggering of charmed hadron production.
We then describe an analysis measuring the prompt charm cross sections at \mbox{\sqs = 7~\TeV}, including a measurement of the \Lcp cross section.
We describe a search for the doubly charmed baryon \Xiccp which was conducted with 0.65~\invpb of 2011 \lhcb data.
We conclude by remarking on future analyses of charmed baryons that are planned at \lhcb.

\section{The \lhcb detector}
The \lhcb~\cite{Alves:2008zz} detector is a dedicated forward arm spectrometer and is the dedicated heavy-flavour physics experiment at the \lhc.
The detector has gathered large volumes of data from 2010 to 2012.
The operating conditions of \lhcb are given in \Tabref{tab:det:beam} along with the original design specifications from the \lhcb Letter of Intent~\cite{Dijkstra:1995cha}.
These demonstrate that \lhcb's operating characteristics have significantly exceeded the orginal design, working at twice the instantaneous luminosity and recording events which are four times as complex as the design.
This has resulted in \lhcb being able to record vastly more charm production than originally envisaged, allowing for greater precision on our measurements.
This data has already been utilised to make several important measurements in the charmed baryon sector, with many more new measurements presently being undertaken.
These will be described in subsequent sections.

\begin{table}[hbt]
\centering
\begin{tabular}{l|r|r|r|r} 

Year & \sqs [\tev] & Instantaneous \lum [$\cm^{-2}s^{-1}$] & \Pmu & \lumi \\ \hline

	2010
	& 7
	& $1\times10^{32}$
	& 0.5 -- 2.5
	& 37 \invpb \\
	
	2011
	& 7
	& $4\times10^{32}$
	& 1.5 
	& 1.0 \invfb \\
	
	2012
	& 8
	& $4\times10^{32}$
	& 1.6 
	& 2 \invfb \\
	
	Design
	& 14
	& $2\times10^{32}$
	& 0.4 
	& - \\
	
	\hline

\end{tabular}

\caption[\lhcb running conditions in the lhc \runone]{
The \lhcb running conditions throughout the \lhc \runone and the nominal conditions.
The pileup, \Pmu, is the average number of \proton~--~\proton collisions in each visible bunch crossing.
The figures provided for the integrated luminosity, \lumi, are the integrated luminosity gathered in that year.
\label{tab:det:beam}
}
\end{table}


The detector possesses powerful resolution of secondary vertices from the decays of heavy-flavour particles produced in the primary interaction.
Heavy flavour particles typically live long enough to fly around 1~\cm from the location of the primary interaction.
\lhcb exploits a high-quality vertex resolution to isolate these particles from the production of lighter hadrons.
The tracking system at \lhcb provides a lifetime resolution of the order of 50~\fs.
The discrimination of interesting signals from the high backgrounds present in a hadronic production environment requires a precise mass resolution and therefore a precise momentum resolution.
The tracking system at \lhcb provides a momentum resolution of $\delta p / p \approx$ 0.4~--~0.6~\%.

Many decays of heavy-flavour hadrons of particular interest have a variety of topologically identical final states, obeying different CP symmetries.
The discrimination between different species of charged particle is therefore of utmost importance.
The particle identification (PID) system at \lhcb is able to provide strong mass hypotheses over the momentum range 1~--~100~\gev.


\section{The \lhcb charm triggers}
The production rate of \ccbar in $\proton\proton$ collisions at the \lhc is much higher than it is in $e^+e^-$ collisions at the \B-factories. This has been measured at \lhcb to be
\begin{equation}
\sigma(\ccbar)_{\pT < 8 \gevc, 2.0 < y < 4.5} = 1419 \pm 12 \stat \pm 116 \syst \pm 65 (\mathrm{fragmentation}) \mub
\end{equation}
in $\proton\proton$ collisions at \mbox{\sqs = 7~\TeV} for production below 8~\gevc transverse momentum and in the rapidity region $\mbox{2.0 -- 4.5}$.
The last error given is the uncertainty from the input fragmentation functions~\cite{LHCb-PAPER-2012-041}.
In \runone, 40~\% of the 5~\khz trigger output rate was allotted to triggers for charmed hadrons.
This presented a difficult challenge to retain as many interesting decays as possible given the constraints on trigger retention.

\begin{figure}[htb]
\centering
\includegraphics[height=3.5in]{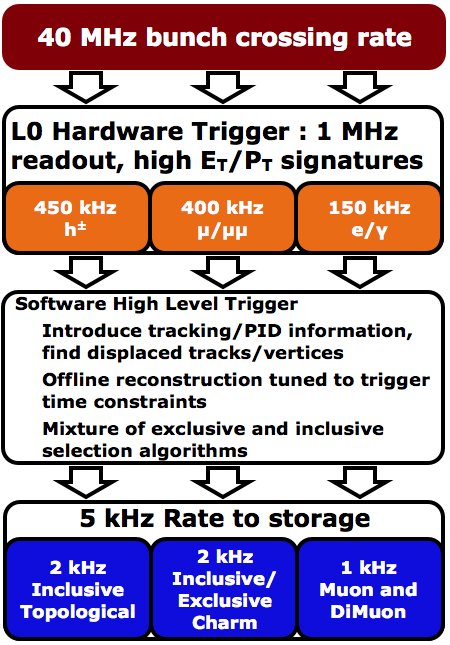}
\caption{The \lhcb trigger schema in \runone.
}
\label{fig:trig}
\end{figure}

The \lhcb trigger is comprised of a low level hardware trigger and a high level software trigger.
This schema is illustrated in \Figref{fig:trig}.
Specifically for hadrons, a cluster must be recorded in the calorimeter exceeding 3.5~\gev.
The first level of the software trigger, the \hltone, exploits a partial reconstruction to quickly and efficiently identify tracks that are displaced from the primary proton-proton interaction point.
For charmed baryon decays to hadronic states a single charged track is required to fulfil a series of track quality cuts.
This track must also have a transverse momentum greater than 1.7~\gevc and an impact parameter with respect to the reconstructed primary interaction greater than 0.1~\mm. 
At the second level of the software trigger, the \hlttwo, a full event reconstruction is employed, allowing the identification of displaced secondary vertices.

The PID information from the \rich detectors is also available at this stage.
For decays of the short lived \Lcp, where the candidate decay vertex is less displaced from the primary proton-proton interaction point than in \D hadron decays, this was of key importance to reject combinations of unrelated tracks.
The dedicated \LcpTopKmpip trigger in the \hlt was one of a limited number of trigger lines which used PID information, to reject candidates where the proton track is a misidentified hadron of another species.
Following this success, direct PID information in the \hlt has now been implemented in a much wider variety of charm exclusive triggers for \runtwo.

Prompt charm is triggered with a series of exclusive lines which exploit hadronic signatures.
A large number of secondary charm from \bquark-hadron decays is also collected with a suite of inclusive \bquark-hadron triggers.
Notably, semileptonic \bquark-hadron decays with muons in the final state can be used to exploit the high efficiency, high purity muonic triggers.
The high production rate of charm necessitates a variety of requirements to be placed on candidates in order to reduce the trigger retention.
These commonly consist of kinematic cuts on the \cquark-hadron decay products, with vertex and track quality cuts.

\section{\Lcp production cross-section at \sqs~=~7~\TeV}
By measuring production cross-sections of charmed hadrons it is possible to test the predictions of quantum chromodynamics (QCD), specifically fragmentation and hadronisation models.
Calculations of charm cross-sections at next-to-leading order made with the Generalized	Mass Variable Flavour Number Scheme (GMVFNS)~\cite{PhysRevD.71.014018, Amsler20081} and at fixed order with next-to-leading-log resummation (FONLL)~\cite{Cacciari:1998it, Cacciari:2003zu, Cacciari:2012ny} have been shown to accurately reproduce the cross-sections measured both at the Tevatron~\cite{PhysRevLett.91.241804} and the \lhc~\cite{ccc1, ccc2} in the central region ($\lvert {\Peta \rvert < 0.5}$) .
\lhcb offers a unique opportunity to study charm production in the forward region in $pp$ collisions.

\lhcb has published a measurement of the cross-sections of the $D^0$, $D^+$, $D_\squark^+$, $D^{*+}$ and \Lcp at a center-of-mass energy \mbox{\sqs = 7 \TeV} which was performed with 15~\invnb of $pp$ collisions recorded in 2010~\cite{LHCb-PAPER-2012-041}.
Charm production at \lhcb can be produced in a variety of ways - directly at the primary interaction, from feed-down of instantaneous decays of excited charm, and from secondary decays of \bquark-hadrons.
Those charmed hadrons produced in the first two ways are denoted as ``prompt'', while those in the latter are denoted as ``secondary'' and are treated as a background in this analysis.

\begin{figure}[htb]
\centering
\subfloat[\Lcp mass]{\includegraphics[width=0.5\textwidth]{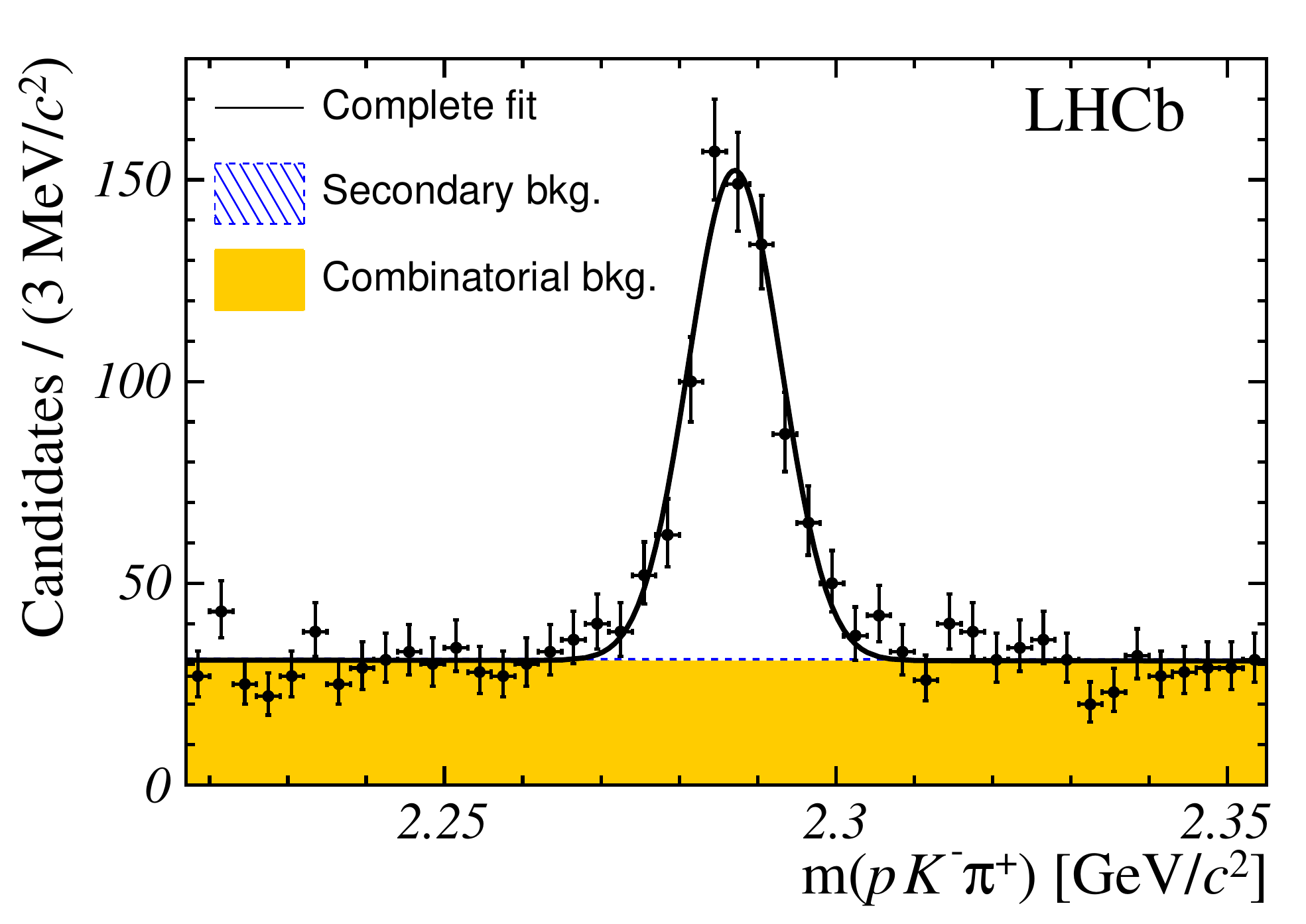}}
\subfloat[\Lcp \logipchisq]{\includegraphics[width=0.5\textwidth]{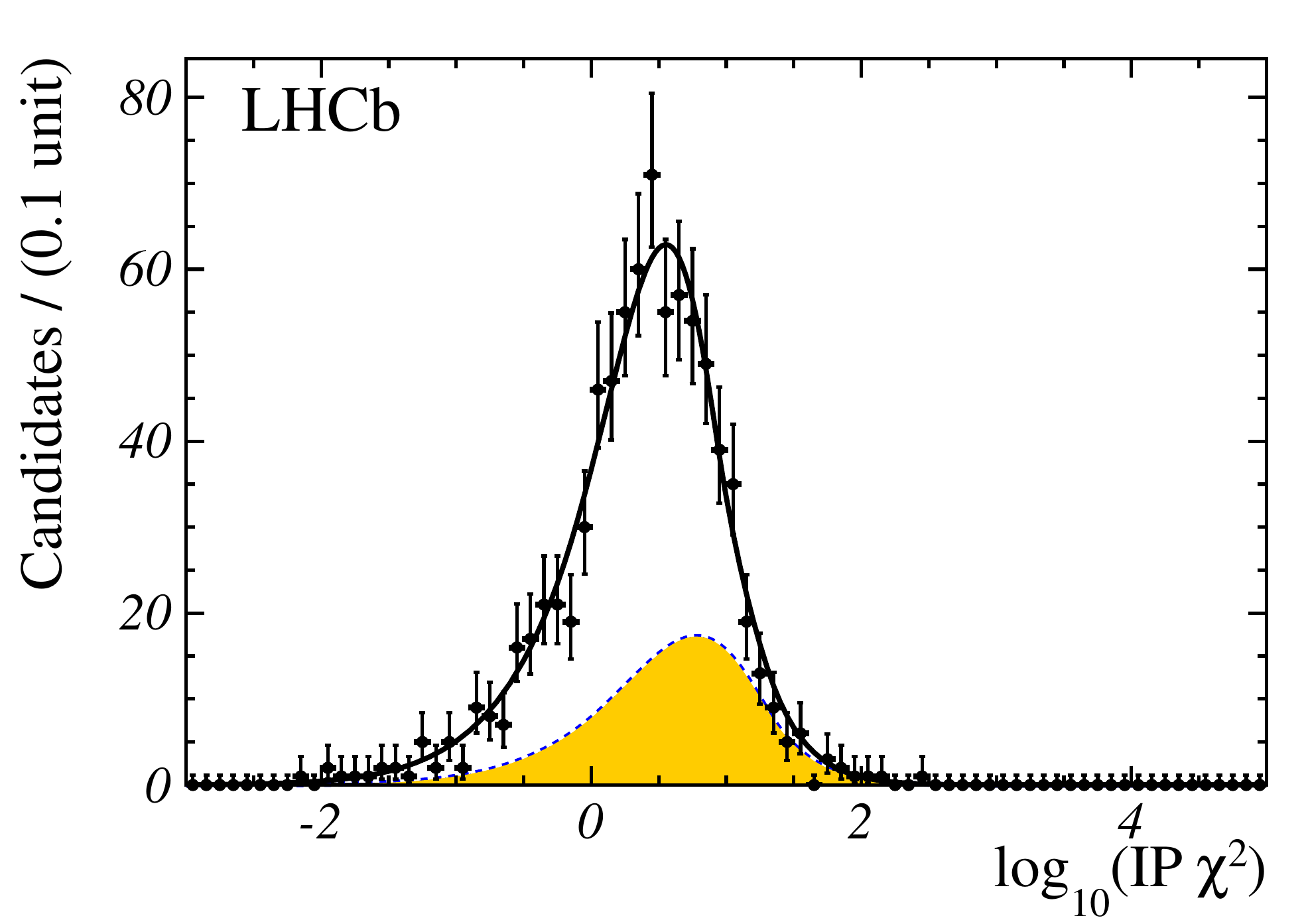}}
\caption{The fits used in the signal extraction of prompt \Lcp from the combinatoric and secondary backgrounds.}
\label{fig:lcsig}
\end{figure}

Candidate \Lcp are selected with kinematic and PID criteria, and data-driven methods and simulation studies are used to evaluate the efficiency of the selection criteria.
The signal extraction is performed with a simultaneous maximum likelihood fit to the \Lcp mass and the \Lcp \logipchisq.
The fits are capable of discriminating among the prompt signal, the secondary background, and the backgrounds arising from random combinations of unrelated tracks.
The results of the \Lcp fit are shown in \Figref{fig:lcsig}, where the secondary contamination was found to be consistent with zero.

\begin{figure}[htb]
\centering
\includegraphics[height=3.1in]{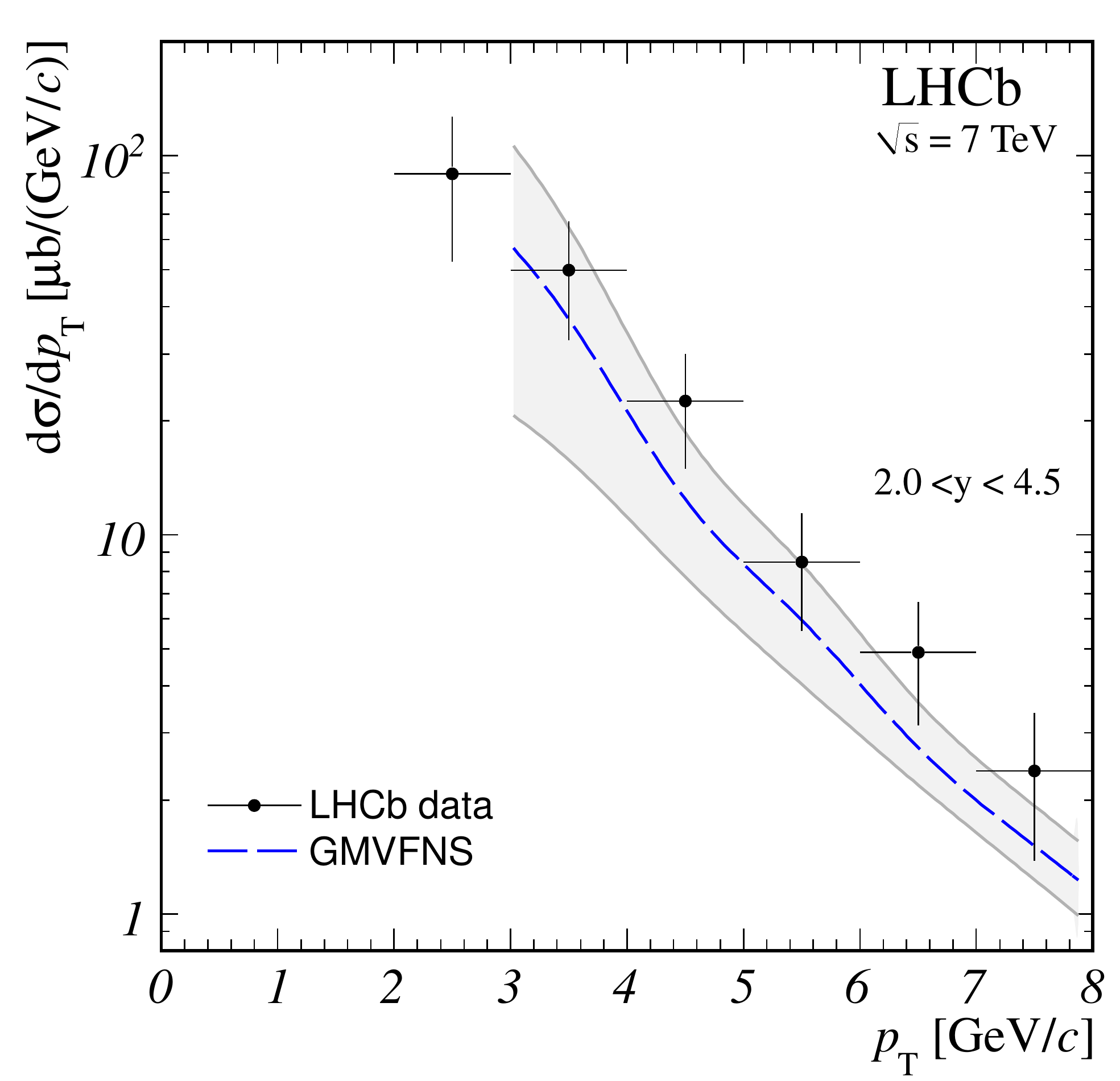}
\caption{The measured \Lcp cross-section as a function of the \Lcp \pT, compared with predictions from the GMVFNS scheme. The error bars show the combined statistical and systematic uncertainty on the measurements and the shaded region indicates the uncertainty on the theoretical predictions.
}
\label{fig:lccs}
\end{figure}

The fiducial region of the measurement was \mbox{2--8~\gevc} transverse momentum and rapidity $\mbox{2.0 -- 4.5}$.
The \Lcp data was divided into bins of \pT with one bin of rapidity, to allow comparison with the predictions from the GMVFNS scheme.
The differential cross-section for \Lcp in \pT bin \Pi is calculated as 
\begin{equation}
\frac{\diff\sigma_i(\Lcp)}{\diff\pT} = \frac{1}{\Delta\pT} \cdot \frac{N_i(\LcpTopKmpip + \mathrm{c.c.})}{\epsilon_i(\LcpTopKmpip) \cdot \BF(\LcpTopKmpip) \cdot \lumi }
\end{equation}
where $N_i(\LcpTopKmpip + \mathrm{c.c.})$ is the number of \LcpTopKmpip decays (and the charge conjugate mode) recorded in the bin, $\epsilon_i(\LcpTopKmpip)$ is the selection efficiency in that bin, $\BF(\LcpTopKmpip)$ is the branching fraction of the decay (taken from the PDG as $\BF(\LcpTopKmpip) = (5.0 \pm 1.3)\%$~\cite{PDG2012}), and \lumi is the integrated luminosity of the data sample.
The \Lcp cross-section as a function of \pT is shown in \Figref{fig:lccs}.

The total \Lcp production cross-section in the fiducial region is calculated by discounting bins with a relative uncertainty larger than 50~\% and extrapolating the cross-section with predictions from \pythia 6.4.
We measure a total cross-section in the acceptance for the \Lcp of
\begin{equation}
\sigma(\Lcp)_{\pT < 8 \gevc, 2.0 < y < 4.5} = 233 \pm 26 \stat \pm 71 \syst \pm 14 (\mathrm{extrapolation}) \mub
\end{equation}


\section{Search for \Xiccp with 2011 data}
For the \sufour multiplets of baryons composed of the \uquark, \dquark, \squark and \cquark quarks we find states in the quark model corresponding to $ccd$ (\Xiccp), $ccu$ (\Xiccpp) and $ccs$ (\Omegacc).
Theoretical calculations generally agree that the \Xiccp state should have lifetimes between 100~--~250~\fs, with the \Xiccpp state having a lifetime between \mbox{500 -- 1550~\fs}~\cite{Guberina:1999mx, Chang:2007xa}.
Evaluations for the particle mass generally predict that the $\Xi_{cc}$ isodoublet will have a mass between \mbox{3.5 -- 3.7~\gevcc}, with the mass of the \Omegacc predicted to be between \mbox{3650 -- 3800~\gevcc}~\cite{Wang:2010hs, Ebert:2002ig, He:2004px, Roberts:2007ni}.

The SELEX collaboration has reported observation of the \Xiccp in its decay to \mbox{\Lc\Km\pip}~\cite{Mattson:2002vu} and evidence of the \Xiccp in the mode \mbox{\proton \Dp \Km}~\cite{Ocherashvili:2004hi}.
The reported state has a measured mass of \mbox{3519~\Ppm~2~\mevcc}, consistent with predictions from the theory community.
Its measured lifetime, however, was consistent with zero, and less than 33~\fs at the 90~\% confidence level.
This is in strong disagreement with predictions of HQET and lattice QCD and is very different to the well-established lifetime of the \Lcp (\mbox{\Ptau $\approx$ 200~\fs}).

If such an observation is legitimate, then much can be learned from the study of the baryon and the mechanisms that lead to its uncharicteristically short lifetime.
Subsequent searches at notably BELLE~\cite{Chistov:2006zj}, BaBar~\cite{Aubert:2006qw} and other experiments have not shown evidence for doubly-charmed baryon production.
As such, the matter is still very much open to discussion, and the theory community eagerly awaits a second experimental observation of the state.

\lhcb has conducted the first search for doubly charmed baryon production at a $pp$ collider~\cite{LHCb-PAPER-2013-049}.
This search was performed on 0.65~\invpb of $pp$ data gathered in 2011 at a centre-of-mass energy \sqs~=~7~\tev, using the decay mode \decay{\Xiccp}{\Lcp(\proton\Km\pip)\Km\pip} and the normalisation channel \LcpTopKmpip.
We specifically measure the ratio of \Xiccp production relative to \Lcp production:
\begin{equation}
R \equiv \frac{\sigma(\Xi_{cc}^{+})\mathcal{B}(\Xi_{cc}^{+} \to \Lambda_{c}^{+}K^{-}\pi^{+})}{\sigma(\Lambda_{c}^{+})} = \frac{N_{\mathrm{sig}}}{N_{\mathrm{norm}}}\frac{\epsilon_{\mathrm{norm}}}{\epsilon_{\mathrm{sig}}}
\end{equation}
where $\sigma$ and $\mathcal{B}$ are the relevant cross-sections and branching fractions, $N_{\mathrm{sig}}$ and $N_{\mathrm{norm}}$ are the extracted yields of the \Xiccp signal and the control \Lcp, and $\epsilon_{\mathrm{sig}}$ and $\epsilon_{\mathrm{norm}}$ are selection efficiencies.

To account for the unknown \Xiccp mass and lifetime we search for the \Xiccp in a wide mass range (\mbox{$\mathrm{3300-3800~MeV/c^{2}}$}), and calculate  efficiencies for a variety of \Xiccp lifetime hypotheses. For each candidate we define a mass difference \dm as
\begin{equation}
\delta m \equiv m([\proton\Km\pip]_{\Lambda_{c}^{+}}\Km\pip) - m([\proton\Km\pip]_{\Lambda_{c}^{+}}) - m(K^{-}) - m(\pi^{+}) 
\end{equation}
where $m([pK^{-}\pi^{-}]_{\Lambda_{c}^{+}}K^{-}\pi^{+})$ is the measured mass of the reconstructed \Xiccp candidate, $m([pK^{-}\pi^{-}]_{\Lambda_{c}^{+}})$ is the measured mass of the reconstructed \Lcp candidate and $m(\Km)$ and $m(\pip)$ are the charged kaon and pion world-averaged masses. The \Xiccp mass window in the analysis corresponds to a $\delta m$ signal window of $380 < \delta m < 880$ \mev. 

The selection of candidates aims to reject backgrounds arising from combinations of unrelated tracks, mis-reconstructed \cquark-hadron and \bquark-hadron decays, and combinations of real \Lcp with unrelated tracks.
We first construct \Lcp candidates, requiring that each passes a selection algorithm in the \hlt that requires that the \Lcp must be displaced from the primary interaction.
The algorithm also places PID, kinematic and vertex/track quality requirements on the decay.
\Xiccp candidates are then constructed by pairing the \Lcp candidates with a kaon and pion track.
The bachelor kaons and pions are required not to point back to the primary interaction.
This diminishes the sensitivity of the selection in the case of a very short lifetime \Xiccp, but is necessary for the rejection of backgrounds where a real \Lcp is paired with random kaons and pions.
Finally, an artificial neural network is used to improve the selection purity of \Xiccp candidates, trained to have as little as possible sensitivity to the \Xiccp lifetime.

The signal yield of the normalisation channel is extracted with a fit to the $m(\proton\Km\pip)$ distribution.
The normalisation yield was found to be $(818 \pm 7) \times 10^{3}$, with a signal width of 6~\MeVcc.
The \Xiccp yield is extracted with an analytic sideband subtraction that requires knowledge of the \Xiccp mass resolution (which is taken from simulation) but requires no other information of the \Xiccp lineshape.

\begin{figure}[htb]
\centering
\includegraphics[height=2.0in]{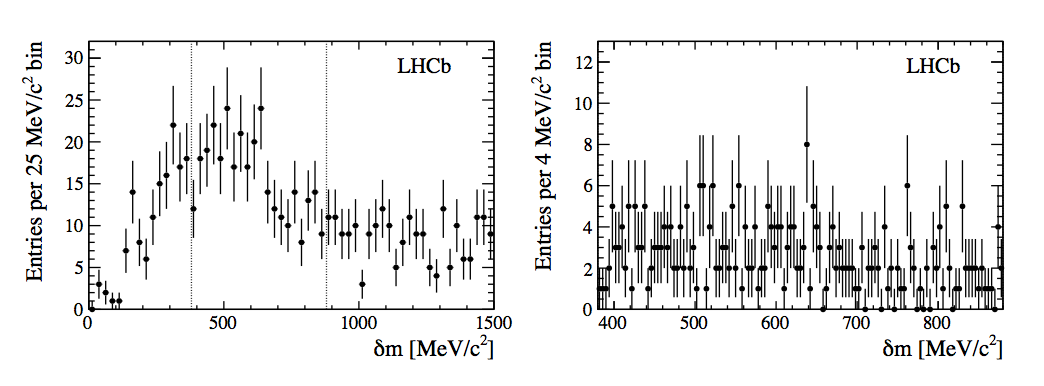}
\caption{The full \dm spectrum (left) and the \Xiccp signal region (right, marked with dashed lines on the full \dm spectrum).
}
\label{fig:dm}
\end{figure}

Signal yields are calculated in 1~\MeVcc intervals of \dm across the full signal region. Local significances at each interval are given as:
\begin{equation}
\mathcal{S} (\delta m) \equiv \frac{N_{S+B}-N_{B}}{\sqrt{\sigma^{2}_{S+B} + \sigma^{2}_{B}}}
\end{equation}
where $\sigma^{2}_{S+B}$ and $\sigma^{2}_{B}$ are the statistical uncertainties on the signal yield and the expected background.
Global significances for each \dm must take into account the ``look elsewhere effect''~\cite{lyons2008}.
We do so by generating a large number of background only pseudo-experiments, with the full analysis method applied to each.
We give a global \Pp-value for a given \PS as the fraction of the total simulated experiments with an equal or larger local significance anywhere in the \dm spectrum.

\begin{figure}[htb]
\centering
\includegraphics[height=3.0in]{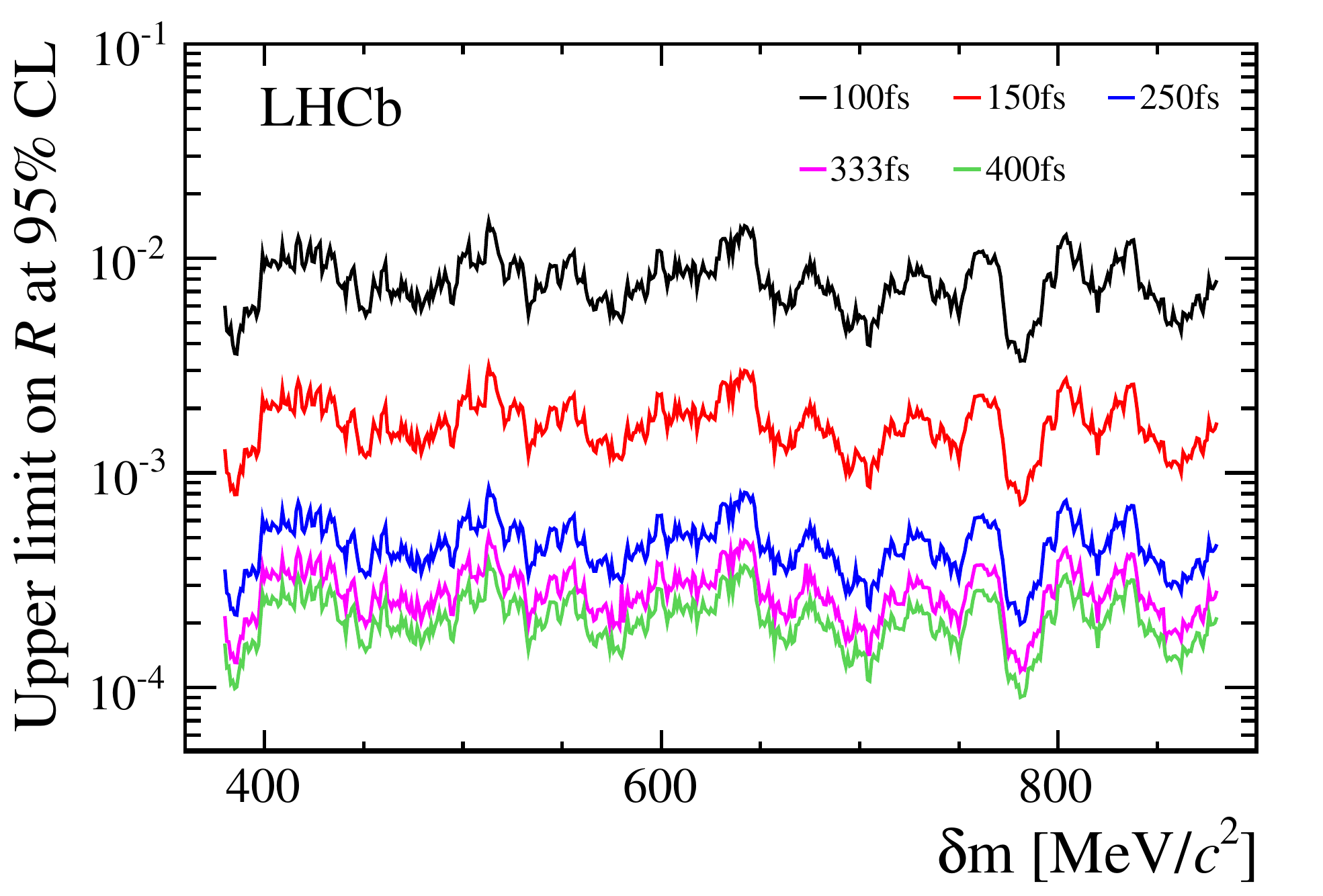}
\caption{Upper limits on $R$ for a number of \Xiccp lifetime hypotheses.}
\label{fig:uls}
\end{figure}

The largest local significance observed is $S = 1.5\sigma$ at a $\dm = 513 \MeVcc$, corresponding to a global \Pp-value for the null hypothesis of 99~\%.
We therefore give upper limits on \PR as a function of \dm with the $CL_S$ method~\cite{Read:2002hq}.
We do so for a variety of \Xiccp lifetime hypotheses (arrived at by re-weighting the simulation-derived efficiencies with different generated \Xiccp lifetimes).
These are given in \Figref{fig:uls}.

\section{Future and ongoing projects}
A great many charmed baryon analyses are currently underway at \lhcb, covering a wide range of topics.
We briefly describe some of these, including updates to the analyses presented herein.
We also detail some future prospects for future \lhcb datasets which will be collected in the \lhc \runtwo and \runthree.

Charm cross-section measurements at \sqs = 13~\tev are in progress with the 2015 \lhcb dataset.
We plan an expanded suite of measurements, including differential cross-sections of the \Lcp, \Sigmacz, \Sigmacpp, \Xicp and \Xicz baryons.
Work is also ongoing in improving the description of the proton detection asymmetry at \lhcb, which is a limiting factor in prospective production asymmetry studies of these baryons.
We hope to perform these measurements with \sqs = 13~\tev data, complementary to our published results on $D$ and $D_{\squark}$ production asymetries at \lhcb.

A new search for the doubly charmed baryons with the full \runone dataset is underway.
In addition to the \Xiccp, we now also search for the \Xiccpp.
The analysis is able to take advantage of improvements to the dedicated \LcpTopKmpip triggers made in 2012 to enhance sensitivity.
An expanded suite of Cabibbo-favoured decay modes will be used to search for these particles, including $\Xi_{cc}^{+[+]} \to D^+(\Km\pip\pip)\proton\Km[\pip]$ and $\Xi_{cc}^{+[+]} \to D^0(\Km\pip)\proton\Km[\pip]$ which can take advantage of \lhcb's excellent $D$ reconstruction.
We tentatively expect to improve our current upper limits on \PR by an order of mangitude.

In addition to these powerful tests of the standard model, there exist promising avenues of investigating new physics with charmed baryons.
Investigations of \CP-violation and of rare decays with charmed baryons offer searches for new physics complementary to those conducted with charmed mesons due to the different strong phases in their decays.
A search for \CP-violation in Cabibbo-suppressed \Lcp decays is currently underway using the \runone dataset.
A search for the rare decay \decay{\Lcp}{\proton\mup\mun} is also being conducted.
In \runtwo we anticipate that charmed baryon \CP-violation searches will have the same sensitivity as our \CP-violation searches with charmed mesons from \runone.
The suite of charmed baryon trigger lines has also been expanded for \runtwo to enable a wide variety of spectroscopy analyses.
We also anticipate that observations of the \Xiccp should be possible at \lhcb with the \runtwo dataset.
Looking ahead to \runthree we anticipate excellent precision on \CP-violation searches with charmed baryons. We also expect the ability to conduct double charm spectroscopy, and the ability to probe for the triple charmed baryon \Omegaccc.

We conclude by reiterating that the \lhcb detector's ability to perform charmed baryon measurements has thus far been excellent.
Over the course of the experiment's lifetime we anticipate a wide number of measurements of charmed baryon branching fractions, masses and lifetimes.
We also plan to search for the remaining charmed baryon states and are optimistic we can contribute to the open questions of doubly charmed baryon production.


\Acknowledgements
We express our gratitude to our colleagues in the CERN
accelerator departments for the excellent performance of the LHC. We
thank the technical and administrative staff at the LHCb
institutes. We acknowledge support from CERN and from the national
agencies: CAPES, CNPq, FAPERJ and FINEP (Brazil); NSFC (China);
CNRS/IN2P3 and Region Auvergne (France); BMBF, DFG, HGF and MPG
(Germany); SFI (Ireland); INFN (Italy); FOM and NWO (The Netherlands);
SCSR (Poland); MEN/IFA (Romania); MinES, Rosatom, RFBR and NRC
``Kurchatov Institute'' (Russia); MinECo, XuntaGal and GENCAT (Spain);
SNSF and SER (Switzerland); NAS Ukraine (Ukraine); STFC (United
Kingdom); NSF (USA). We also acknowledge the support received from the
ERC under FP7. The Tier1 computing centres are supported by IN2P3
(France), KIT and BMBF (Germany), INFN (Italy), NWO and SURF (The
Netherlands), PIC (Spain), GridPP (United Kingdom). We are thankful
for the computing resources put at our disposal by
Yandex LLC (Russia), as well as to the communities behind the multiple open
source software packages that we depend on.

\setboolean{inbibliography}{true}
\bibliographystyle{unsrt}
\bibliography{bib/main,bib/LHCb-PAPER,bib/LHCb-CONF,bib/LHCb-DP,bib/lambdacexp,bib/theory,bib/pid,bib/detector}

\begin{thebibliography}{10}

\bibitem{Chistov:2006zj}
R.~Chistov et~al.
\newblock {Observation of new states decaying into \mbox{$\Lambda_c^+ K^-
  \pi^+$} and \mbox{$\Lambda_c^+ K_0^S \pi^-$}}.
\newblock {\em Phys.Rev.Lett.}, 97:162001, 2006.

\bibitem{Lesiak:2006sk}
Tadeusz Lesiak.
\newblock {Charmed baryon spectroscopy with Belle}.
\newblock 2006.

\bibitem{Mizuk:2004yu}
R.~Mizuk et~al.
\newblock {Observation of an isotriplet of excited charmed baryons decaying to
  \mbox{$\Lambda_c^+ \pi$}}.
\newblock {\em Phys.Rev.Lett.}, 94:122002, 2005.

\bibitem{Aubert:2007dt}
Bernard Aubert et~al.
\newblock {A Study of Excited Charm-Strange Baryons with Evidence for new
  Baryons $\Xi_c^{+}(3055)$ and $\Xi_c^{+}(3123)$}.
\newblock {\em Phys.Rev.}, D77:012002, 2008.

\bibitem{Lesiak:2008wz}
T.~Lesiak et~al.
\newblock {Measurement of masses of the $\Xi_c(2645)$ and $\Xi_c(2815)$ baryons
  and observation of \mbox{$\Xi_c(2980) \to \Xi_c(2645)\pi$}}.
\newblock {\em Phys.Lett.}, B665:9--15, 2008.

\bibitem{Aubert:2007bt}
Bernard Aubert et~al.
\newblock {Production and decay of $\Omega_{c}^{0}$}.
\newblock {\em Phys.Rev.Lett.}, 99:062001, 2007.

\bibitem{Aubert:2006je}
Bernard Aubert et~al.
\newblock {Observation of an excited charm baryon $\Omega_c^{*}$ decaying to
  \mbox{$\Omega_c0 \gamma$}}.
\newblock {\em Phys.Rev.Lett.}, 97:232001, 2006.

\bibitem{Solovieva:2008fw}
E.~Solovieva, R.~Chistov, I.~Adachi, H.~Aihara, K.~Arinstein, et~al.
\newblock {Study of $\Omega_c^0$ and $\Omega_c^{*0}$ Baryons at Belle}.
\newblock {\em Phys.Lett.}, B672:1--5, 2009.

\bibitem{PDG2012}
J.~Beringer et~al.
\newblock Review of particle physics.
\newblock {\em Phys. Rev.}, D86:010001, 2012.

\bibitem{Alves:2008zz}
A.~A. Alves~Jr. et~al.
\newblock {The \lhcb detector at the LHC}.
\newblock {\em JINST}, 3:S08005, 2008.

\bibitem{Dijkstra:1995cha}
H.~Dijkstra, Hans~Jürgen Hilke, Tatsuya Nakada, and Thomas Ypsilantis.
\newblock {LHCb Letter of Intent, LHCb Collaboration}.
\newblock 1995.

\bibitem{LHCb-PAPER-2012-041}
R.~Aaij et~al.
\newblock {Prompt charm production in $pp$ collisions at $\sqrt{s}=7$ TeV}.
\newblock {\em Nucl. Phys.}, B871:1, 2013.

\bibitem{PhysRevD.71.014018}
B.~A. Kniehl, G.~Kramer, I.~Schienbein, and H.~Spiesberger.
\newblock Inclusive ${D}^{*\ifmmode\pm\else\textpm\fi{}}$ production in
  $p\overline{p}$ collisions with massive charm quarks.
\newblock {\em Phys. Rev. D}, 71:014018, Jan 2005.

\bibitem{Amsler20081}
C.~Amsler et~al.
\newblock Review of particle physics.
\newblock {\em Physics Letters B}, 667(1–5):1 -- 6, 2008.
\newblock Review of Particle Physics.

\bibitem{Cacciari:1998it}
Matteo Cacciari, Mario Greco, and Paolo Nason.
\newblock {The P(T) spectrum in heavy flavor hadroproduction}.
\newblock {\em JHEP}, 05:007, 1998.

\bibitem{Cacciari:2003zu}
Matteo Cacciari and Paolo Nason.
\newblock {Charm cross-sections for the Tevatron Run II}.
\newblock {\em JHEP}, 09:006, 2003.

\bibitem{Cacciari:2012ny}
Matteo Cacciari, Stefano Frixione, Nicolas Houdeau, Michelangelo~L. Mangano,
  Paolo Nason, and Giovanni Ridolfi.
\newblock {Theoretical predictions for charm and bottom production at the LHC}.
\newblock {\em JHEP}, 10:137, 2012.

\bibitem{PhysRevLett.91.241804}
D.~Acosta et~al.
\newblock Measurement of prompt charm meson production cross sections in
  $p\overline{p}$ collisions at $\sqrt{s}=1.96\text{ }\text{
  }\mathrm{T}\mathrm{e}\mathrm{V}$.
\newblock {\em Phys. Rev. Lett.}, 91:241804, Dec 2003.

\bibitem{ccc1}
B.~Abelev et~al.
\newblock Measurement of charm production at central rapidity in proton-proton
  collisions at $\sqrt {s} = 2.76\;{\text{tev}}$.
\newblock {\em Journal of High Energy Physics}, 2012(7), 2012.

\bibitem{ccc2}
B.~Abelev et~al.
\newblock Measurement of charm production at central rapidity in proton-proton
  collisions at $\sqrt {s} = 7{ }tev$.
\newblock {\em Journal of High Energy Physics}, 2012(1), 2012.

\bibitem{Guberina:1999mx}
B.~Guberina, B.~Melic, and H.~Stefancic.
\newblock {Inclusive decays and lifetimes of doubly charmed baryons}.
\newblock {\em Eur.Phys.J.}, C9:213--219, 1999.

\bibitem{Chang:2007xa}
Chao-Hsi Chang, Tong Li, Xue-Qian Li, and Yu-Ming Wang.
\newblock {Lifetime of doubly charmed baryons}.
\newblock {\em Commun.Theor.Phys.}, 49:993--1000, 2008.

\bibitem{Wang:2010hs}
Zhi-Gang Wang.
\newblock {Analysis of the ${1\over 2}^+$ doubly heavy baryon states with QCD
  sum rules}.
\newblock {\em Eur.Phys.J.}, A45:267--274, 2010.

\bibitem{Ebert:2002ig}
D.~Ebert, R.N. Faustov, V.O. Galkin, and A.P. Martynenko.
\newblock {Mass spectra of doubly heavy baryons in the relativistic quark
  model}.
\newblock {\em Phys.Rev.}, D66:014008, 2002.

\bibitem{He:2004px}
Da-Heng He, Ke~Qian, Yi-Bing Ding, Xue-Qian Li, and Peng-Nian Shen.
\newblock {Evaluation of spectra of baryons containing two heavy quarks in bag
  model}.
\newblock {\em Phys.Rev.}, D70:094004, 2004.

\bibitem{Roberts:2007ni}
W.~Roberts and Muslema Pervin.
\newblock {Heavy baryons in a quark model}.
\newblock {\em Int.J.Mod.Phys.}, A23:2817--2860, 2008.

\bibitem{Mattson:2002vu}
M.~Mattson et~al.
\newblock {First observation of the doubly charmed baryon $\Xi_{cc}^{+}$}.
\newblock {\em Phys.Rev.Lett.}, 89:112001, 2002.

\bibitem{Ocherashvili:2004hi}
A.~Ocherashvili et~al.
\newblock {Confirmation of the double charm baryon $\Xi_{cc}^{+}(3520)$ via its
  decay to \mbox{$p D^+ K^-$}}.
\newblock {\em Phys.Lett.}, B628:18--24, 2005.

\bibitem{Aubert:2006qw}
Bernard Aubert et~al.
\newblock {Search for doubly charmed baryons $\Xi_{cc}^+$ and $\Xi_{cc}^{++}$
  in BABAR}.
\newblock {\em Phys.Rev.}, D74:011103, 2006.

\bibitem{LHCb-PAPER-2013-049}
R.~Aaij et~al.
\newblock {Search for the doubly charmed baryon $\Xi_{cc}^+$}.
\newblock {\em JHEP}, 12:090, 2013.

\bibitem{lyons2008}
Louis Lyons.
\newblock Open statistical issues in particle physics.
\newblock {\em Ann. Appl. Stat.}, 2(3):887--915, 09 2008.

\bibitem{Read:2002hq}
Alexander~L. Read.
\newblock {Presentation of search results: The CL(s) technique}.
\newblock {\em J. Phys.}, G28:2693--2704, 2002.
\newblock [,11(2002)].

\end{thebibliography}

\end{document}